\documentclass{eptcs}
\providecommand{\event}{TTC 2013}

\usepackage{oz2}
\usepackage{graphicx}
\usepackage{alltt}
\usepackage{breakurl}

\title{Solving the Petri-Nets to Statecharts Transformation Case with UML-RSDS}
\author{K. Lano, S. Kolahdouz-Rahimi, K. Maroukian\\
Dept. of Informatics, King's College London, Strand, London, UK\thanks{Research supported by the HoRTMoDA EPSRC project}}

\def\titlerunning{Petri-Nets to Statecharts Case with UML-RSDS}
\def\authorrunning{K. Lano, S. Kolahdouz-Rahimi, K. Maroukian}

\begin{document}
\maketitle

\begin{abstract}
This paper provides a solution to the
Petri-Nets to statecharts case using
UML-RSDS. We show how a highly
declarative solution which is 
confluent and invertible
can be given using this
approach.
\end{abstract}



\section{Introduction}

This case study \cite{GR13}
is an update-in-place
transformation which simultaneously
modifies (by deletion and simplification)
an input Petri-Net model, and (by
construction and elaboration) an output
statechart model. We provide a 
specification of the transformation
in the UML-RSDS language \cite{umlrsds}
and show that this is terminating, 
confluent and invertible.

UML-RSDS is a model-based development
language and
toolset, which specifies systems in a \linebreak
platform-independent manner, and 
provides automated code generation
from these specifications
to executable implementations
(in Java, C$^\#$ and C++). Tools for analysis
and verification are also provided.
Specifications are expressed using the
UML 2 standard language: class diagrams
define data, use cases define the 
top-level services or functions of the
system, and operations can be used to
define detailed functionality. Expressions,
constraints, pre and postconditions and
invariants all use the standard OCL
notation of UML 2.

For model transformations, the class
diagram expresses the metamodels
of the source and target models, and
auxiliary data can also be defined.
Use cases define the main transformation
phases of the transformation: each use
case has a set of pre and postconditions
which define its intended functionality.

The Petri Net to statecharts
transformation can be sequentially
decomposed into three subtransformations: an $initialise$
transformation, which copies the
essential structure of the Petri Net to
an initial statechart, followed by the
main $pn2sc$ reduction/elaboration
transformation. A final $cleanup$
transformation removes elements which
do not contribute to the target 
structure.

\begin{figure}[htbp]
\centering
\includegraphics[width=5.2in]{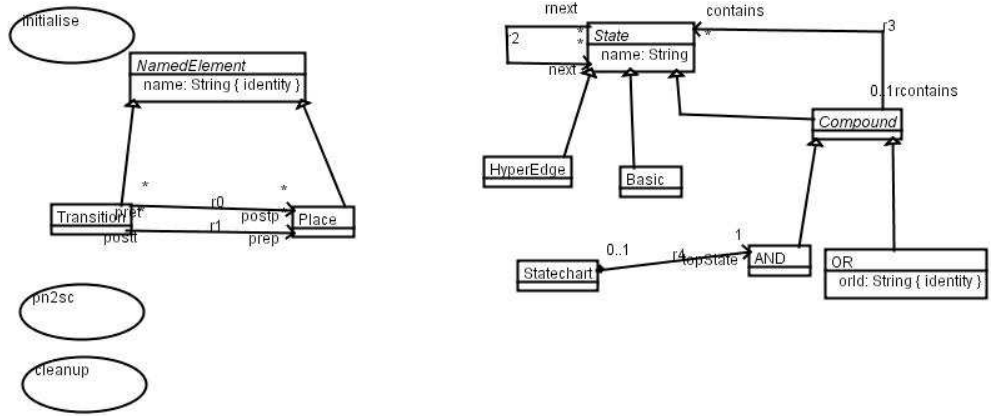}
\caption{PN 2 SC metamodels}
\label{pn2scmm}
\end{figure}

Figure \ref{pn2scmm} shows the source
and target
metamodels of the transformation,
and the three use cases representing
the sub-transformations.

We extend \cite{GR13} by asserting
that $name$ is unique for $HyperEdge$,
$Basic$ and $OR$:
\[ HyperEdge{\fun}isUnique(name) \\
 Basic{\fun}isUnique(name) \\
 OR{\fun}isUnique(name) \]
 This means that object indexing by
name can be used for these entity
types: $OR[s]$ denotes the or-state with
name $s : String$, for example, if 
such a state exists.

 
\section{Initialisation transformation}

This has the precondition that the
statechart is unpopulated: 
$State.size = 0$,
$Statechart.size = 0$, 
and that $name$ is
unique for $NamedElement$s. There
 are four postconditions,
which define the intended state at 
termination of the transformation.
These postconditions are also
interpreted as definitions of the
transformation steps.

Postcondition $I1$ applies to elements
of $Place$ to map them to $Basic$ and
$OR$ states:
\begin{alltt}
\(Basic{\fun}exists( b | b.name = name \& \)
        \(OR{\fun}exists( o | o.name = name \& b : o.contains ) )\)
\end{alltt}
Logically this can be read as 
``for all $p$ in $Place$, there exists $b$ in $Basic$
with $b.name = p.name$, and $o$ in $OR$
with $o.name = p.name$ and 
$b$ in $o.contains$". The inverse link $rcontains$
is set implicitly ($o : b.rcontains$).

Postcondition $I2$ applies to $Transition$s
to map them to $HyperEdge$s:
\begin{alltt}
\(HyperEdge{\fun}exists( e | e.name = name ) \)
\end{alltt}

$I3$ sets up the next/rnext links
between hyperedges and basic states
based upon the corresponding 
postt/prep links in the Petri Net:
\begin{alltt}
\(t : postt  \implies  HyperEdge[t.name] : Basic[name].next \)
\end{alltt}
applied to $Place$ (``if $t$ is a post-transition
of self, then the hyperedge corresponding
to $t$ is in the next states of the basic
state corresponding to self").

$I4$ sets up the next/rnext links
between basic states and hyperedges
based upon the corresponding 
postp/pret links in the Petri Net:
\begin{alltt}
\(p : postp  \implies  Basic[p.name] : HyperEdge[name].next \)
\end{alltt}
applied to $Transition$.

This transformation uses the `Map
objects before links' pattern \cite{Lano12jss} to separate mapping of
elements and their links. It avoids
the need for recursive processing: each
of $I1$, ..., $I4$ can be implemented by
a linear iteration over their source
domains. This implementation is 
generated automatically by UML-RSDS
as a Java program.

Termination, confluence and invertibility
of such transformations follows by
construction \cite{Lano12jss}. The
computational complexity is linear in
$NamedElement.size$. The transformation
establishes \linebreak $Basic{\fun}isUnique(name)$,
$HyperEdge{\fun}isUnique(name)$
and $OR{\fun}isUnique(name)$ because
of the uniqueness of names of named
elements. Indeed these properties are
invariants of $initialise$.


\section{Main transformation}

This has as its preconditions $I1$,
$I2$, $I3$, $I4$, 
together with the uniqueness properties
of $name$ for $Basic$, $HyperEdge$ and
$OR$, and that $AND$ is empty.
An invariant $Inv$ asserts that for all
places, there is a unique OR state with
the same name:
\[ Place{\fun}forAll( p | OR{\fun}exists1( o | o.name = p.name ) ) \]
This ensures that there is an injective
function $equiv: Place \fun OR$.
In our notation, $OR[p.name]$ is
$equiv(p)$ for $p : Place$.

The uniqueness properties 
of $name$ for $Basic$,
$HyperEdge$ and $OR$ are also invariant.
$Inv$ is established by $initialise$ 
because of postcondition $I1$ and the
uniqueness of $name$ on OR.

The highest priority rule (postcondition)
is $Post1$, which performs the 
OR-reduction of \cite{GR13} on
$Transition$ instances:
\begin{alltt}
\(prep.size = 1 \& postp.size = 1 \&\)
\(q : prep \& r : postp \&\)
\((q.pret \cap r.pret){\fun}size() = 0 \&\)
\((q.postt \cap r.postt){\fun}size() = 0  \implies\)
    \(OR{\fun}exists( p | p.name = q.name + ``\_OR\_" + r.name \&\)
    \(      p.contains = OR[q.name].contains \cup OR[r.name].contains \&\)
    \(      q.name = p.name ) \&\)
    \(      q.pret{\fun}includesAll(r.pret) \&\)
    \(      q.postt{\fun}includesAll(r.postt) \&\)
    \(      r{\fun}isDeleted() \& \)
    \(      self{\fun}isDeleted() \)
\end{alltt}
This follows very closely the specification
in \cite{GR13}, with $self : Transition$
playing the role of $t$.
The updates to the Petri-Net are the last
five lines, $q$ replaces the $q \fun self \fun r$ structure and is renamed to match
the new OR state, thus maintaining $Inv$.


For AND-reduction there are two 
postconditions/rules for the symmetric
cases: $Post2$ merges pre-places
with equivalent connectivities, and
again is applied to each $Transition$:
\begin{alltt}
\(p1 : prep \& prep.size > 1 \& \)
\(prep{\fun}forAll( p2 | p1.pret = p2.pret \& p1.postt = p2.postt )  \implies\)
    \(AND{\fun}exists( a | OR{\fun}exists( p |\)
        \(a : p.contains \& a.contains = OR[prep.name] \& \)
        \(p.name = ``AND1\_" + name \& a.name = ``a1\_" + name \&\)
        \(p1.name = ``AND1\_" + name ) ) \&\)
        \((prep - \{p1\}){\fun}isDeleted() \&\)
        \(prep = Set\{p1\}\)
\end{alltt}
The last three lines define the update
to the Petri-Net: all $prep$ places of
$self$ are deleted except for $p1$, which
is renamed to match the newly
created OR state (therefore maintaining
$Inv$).

$Post3$ merges post-places with
equivalent connectivities, for each
applicable $Transition$:
\begin{alltt}
\(p1 : postp \& postp.size > 1 \& \)
\(postp{\fun}forAll( p2 | p1.pret = p2.pret \& p1.postt = p2.postt )  \implies\)
    \(AND{\fun}exists( a | OR{\fun}exists( p |\)
        \(a : p.contains \& a.contains = OR[postp.name] \& \)
        \(p.name = ``AND2\_" + name \& a.name = ``a2\_" + name \&\)
        \(p1.name = ``AND2\_" + name ) ) \&\)
        \((postp - \{p1\}){\fun}isDeleted() \&\)
        \(postp = Set\{p1\}\)
\end{alltt}
This maintains $Inv$ for the same
reason as $Post2$.

\section{Cleanup transformation}

This transformation deletes
OR states with empty contents:
\[ contains.size = 0 ~\implies~ self{\fun}isDeleted() \]
on $OR$.

Finally, an instance $sc : Statechart$
needs to be created, with 
$sc.topState$ being the unique
topmost AND state
produced by the main transformation,
if such a state exists:
\[ v = OR{\fun}select( rcontains.size = 0 ) ~ \& ~ v.size = 1 ~\&~ ox : v ~~\implies\\ 
\t2 AND{\fun}exists( a | a.name = 
``\_TOPSTATE\_" ~\&~ ox : a.contains ) \]
and
\[ w = AND{\fun}select( rcontains.size = 0 ) ~ \& ~ w.size = 1 ~\&~ax : w ~~\implies\\ 
\t4     Statechart{\fun}exists( sc | sc.topState = ax ) \]

This transformation is terminating and
semantically correct by
construction.


\section{Results}

Table \ref{perfres2} gives the test results
for the performance tests for the Java 4 executable 
in the 
SHARE environment, and for the Java 6, C\# and C++ 
executables on a standard Windows 7 laptop. 

\begin{table}[htbp]
\centering
\begin{tabular}{l|l|l|l|l}
{\em Test} & {\em Transformation execution time}: Java 4 & Java 6 & C\# & C++ \\ \hline
sp200 & 100ms & 15ms & 29ms & 0s\\
sp500 & 160ms & 31ms & 63ms & 2s \\
sp1000 & 290ms & 94ms & 198ms & 6s \\
sp5000 & 3815ms & 1670ms & 5069ms & 161s\\
sp10000 & 13713ms & 6614ms & 21980ms & --\\
sp20000 & 48s & 35s & 87s & --\\
sp40000 & 258s & 177s & 468s & -- \\
sp80000 & 3142s & 5619s & 10003s & --  
\end{tabular}
\caption{Performance test results for Java, C\# and C++}
\label{perfres2}
\end{table}

The results for the Java 4, C$\#$
and Java 6 (which uses HashSet instead of
Vector for sets) implementations were
quite similar, which is in contrast to problems 
involving uni-directional associations, where the 
Java 6 translation is typically 100 times more
efficient than the Java 4 version. C++ has efficiency
problems for complex collection manipulations as used in
this case study. 
All the versions may be
found at \url{http://www.dcs.kcl.ac.uk/staff/kcl/uml2web/pn2sc/}. 

Table \ref{summtab} shows the 
summary table completed for our
solution.

\begin{table}[htbp]
\centering
\begin{small}
\begin{tabular}{l|l|l|l|l|l|l|l|l}
{\em Solution} & {\em Language} & {\em Perform.} & 5.2.1: & 5.2.2: & 5.2.3: & 5.2.4: & 5.2.5: & 5.2.6: \\
{\em Name} & {\em (for all} & {\em optimis-} & verifi- & simu- & change- & reverse & debug. & refact- \\
  & {\em aspects)} & {\em ations} & cation & lation & prop & & & oring \\ \hline
UML-RSDS & UML-RSDS & E & CT & N & N & Y & N & N \\
\end{tabular}
\caption{Solution table}
\label{summtab}
\end{small}
\end{table}

The optimisation provided (for rules
$Post1$, $Post2$, $Post3$)
is to omit tests for the
truth of the succedent of the rule
(ie., the negative application condition
of the rule) when
applying the rule: the system can 
detect that a formula such as
$self{\fun}isDeleted()$ is inconsistent
with the positive application condition
of the rule, and therefore that there is
no need to evaluate the formulae 
before applying the rule.

The transformation can be reversed by
reversing the initialisation.

\end{document}